# On the calculation of percentile-based bibliometric indicators

Ludo Waltman[1] and Michael Schreiber[2]

[1] Centre for Science and Technology Studies, Leiden University, Leiden, The Netherlands
waltmanlr@cwts.leidenuniv.nl

[2] Institute of Physics, Chemnitz University of Technology, 09107 Chemnitz, Germany
schreiber@physik.tu-chemnitz.de

A percentile-based bibliometric indicator is an indicator that values publications based on their position within the citation distribution of their field. The most straightforward percentile-based indicator is the proportion of frequently cited publications, for instance the proportion of publications that belong to the top 10% most frequently cited of their field. Recently, more complex percentile-based indicators were proposed. A difficulty in the calculation of percentile-based indicators is caused by the discrete nature of citation distributions combined with the presence of many publications with the same number of citations. We introduce an approach to calculating percentile-based indicators that deals with this difficulty in a more satisfactory way than earlier approaches suggested in the literature. We show in a formal mathematical framework that our approach leads to indicators that do not suffer from biases in favor of or against particular fields of science.

## 1. Introduction

Traditionally, when using citation analysis to assess the impact of the work of for instance a research group, a university, or a journal, the main indicator is the average number of citations per publication, preferably with a normalization that corrects for differences in citation behavior between scientific fields. However, citation distributions tend to be highly skewed, and therefore the average number of citations per publication may be strongly influenced by one or a few publications with a very large number of citations (e.g., Waltman et al., in press, Section 4.1). This is often considered undesirable. For this reason, more and more attention is paid to alternative citation-based impact indicators. One class of alternative indicators is based on the idea of looking at the position of a publication within the citation distribution of its field rather than at the actual number of citations of a publication. The position of a publication within the citation distribution of its field is expressed in terms of a percentile of the citation distribution. In this paper, our focus is on this class of alternative indicators. The indicators are referred to as percentile-based indicators in this paper.

A simple commonly used percentile-based indicator is the proportion of frequently cited publications (Tijssen, Visser, & Van Leeuwen, 2002; Van Leeuwen, Visser, Moed, Nederhof, & Van Raan, 2003). This indicator, which we refer to as the $PP_{top\ x\%}$ indicator, calculates the proportion of the publications of for instance a research group that belong to the top $x\%$ most frequently cited of their field. The focus is often on the top 10% most frequently cited publications of a field (e.g., Bornmann, De Moya Anegón, & Leydesdorff, 2012; Waltman et al., in press), in which case one obtains the $PP_{top\ 10\%}$ indicator. Of course, instead of the top 10%, one could also look at for instance the top 1%, top 2%, top 5%, or top 20% (e.g., Lewison,

Thornicroft, Szmukler, & Tansella, 2007; National Science Board, 2012). Recently, more complex percentile-based indicators have been proposed (e.g., Leydesdorff, Bornmann, Mutz, & Opthof, 2011).

A difficulty in the calculation of percentile-based indicators is caused by the discrete nature of citation distributions. Suppose that in a particular field there are 105 publications: 94 without citations, 1 with 10 citations, and 10 with 20 citations each. Which of these publications belong to the top 10% most frequently cited of the field? Clearly, the 10 publications with 20 citations belong to the top 10%. But what about the publication with 10 citations? If this publication is counted as a top 10% publication, we in fact have 11 / 105 = 10.48% top 10% publications. Conversely, if this publication is not counted as a top 10% publication, we have 10 / 105 = 9.52% top 10% publications. In other words, because of the discrete nature of citation distributions, it is often not possible to exactly define the top 10% most frequently cited publications of a field.

Usually, there are multiple publications in a field that have the same number of citations. The presence of such 'ties' may aggravate the above problem. To see this, consider again a field with 105 publications. This time there are 90 publications without citations, 10 publications with 10 citations each, and 5 publications with 20 citations each. If the publications with 10 citations are counted as top 10% publications, we end up with 15 / 105 = 14.29% top 10% publications. Conversely, if the publications with 10 citations are not counted as top 10% publications, we have only 4.76% top 10% publications. Clearly, both outcomes are unsatisfactory. The problem becomes especially serious when making comparisons between fields. If in one field we have 12% top 10% publications while in another field we have only 8% top 10% publications, the $PP_{top\ 10\%}$ indicator would be strongly biased. A research group active in the former field would be favored over a research group active in the latter field.

Our aim in this paper is to introduce an approach to calculating percentile-based indicators that deals with the discrete nature of citation distributions and the presence of ties in a proper way. The approach that we introduce leads to indicators that do not suffer from biases in favor of or against particular fields. In the case of the $PP_{top\ 10\%}$ indicator, our approach ensures that in each field we have exactly 10% top 10% publications.

The organization of this paper is as follows. First, in Section 2, we briefly illustrate the empirical relevance of the problem that we study. Next, in Section 3, we provide an overview of a number of approaches that have been suggested to calculate percentile-based indicators. In Section 4, we introduce an alternative approach and argue that this approach is preferable over the approaches proposed in earlier research. Finally, in Section 5, we present our alternative approach in a more formal mathematical framework. We note that the discussion in Sections 2, 3, and 4 focuses on one specific percentile-based indicator: The $PP_{top\ 10\%}$ indicator. Section 5 relates to percentile-based indicators in general.

## 2. Some empirical context

To get some insight into the empirical relevance of the problem studied in this paper, let us have a look at some real-world citation distributions. For seven fields (i.e., subject categories) in the Web of Science database, we collected all publications of the document type *article* that appeared in 1999. For each publication, we counted the number of citations received by the end of 2003. Table 1 reports for each field the top 10% threshold, calculated as the 90th percentile of the citation distribution of the field.[1] In addition, the table reports for each field the percentage of publications whose number of citations is below the top 10% threshold, exactly at the threshold, or above the threshold. As can be seen, the percentage of publications below the threshold varies between 86.9% and 90.0%, while the percentage of publications above the threshold ranges from 9.6% to 10.0%. Between 0.4% and 3.6% of the publications in a field are exactly at the threshold. As is to be expected, this percentage is highest in fields with a relatively low citation density. In Table 1, these are the fields of economics and mathematics.

Table 1. Percentage of the publications in a field whose number of citations is below, exactly at, or above the top 10% threshold.

| Field | No. pub. | No. cit. at threshold | % pub. below threshold | % pub. at threshold | % pub. above threshold |
|---|---|---|---|---|---|
| Biochemistry & molecular biology | 42,749 | 39 | 90.0 | 0.4 | 9.6 |
| Cardiac & cardiovascular systems | 10,885 | 29 | 89.6 | 0.5 | 9.8 |
| Chemistry, analytical | 13,675 | 17 | 89.0 | 1.1 | 9.9 |
| Economics | 7,580 | 9 | 88.6 | 1.6 | 9.8 |
| Mathematics | 12,680 | 5 | 86.9 | 3.6 | 9.6 |
| Physics, applied | 26,562 | 14 | 89.1 | 1.0 | 9.9 |
| Surgery | 21,288 | 15 | 88.9 | 1.1 | 10.0 |

## 3. Overview of different approaches to calculating percentile-based indicators

In this section, we provide an overview of a number of approaches to calculating percentile-based indicators. These approaches have been suggested in earlier papers. To illustrate the different approaches, we use an example already introduced above. In this example, there is a field with 105 publications: 90 without citations, 10 with 10 citations each, and 5 with 20 citations each. We focus on the $PP_{top\ 10\%}$ indicator, so our interest is in publications that belong to the top 10% of their field. Clearly, in our example, the 90 publications without citations do not belong to the top 10% of their field, while the 5 publications with 20 citations do belong to the top 10%. The difficulty is in the way in which the 10 publications with 10 citations should be handled. We therefore pay attention mostly to these publications.

We first discuss the approach proposed by Leydesdorff et al. (2011).[2] In this approach, for each publication a corresponding percentile is determined. This is done based on the number of publications with fewer citations than the publication of interest. In our example, there are 90 publications with fewer than 10 citations. The publications with 10 citations therefore have percentile 90 / 105 = 85.7 (or 86, when rounded to an integer, as suggested by Leydesdorff et al.), which means that they are not counted as top 10% publications. Hence, we have only 5 / 105 = 4.76% top 10% publications.

---

[1] The 90th percentile of a citation distribution is not always clearly defined. For instance, in a field with 10 publications, 9 without citations and 1 with 10 citations, it is not immediately clear whether the 90th percentile equals 0 or 10 (or perhaps even 5) citations. However, due to the presence of a lot of ties in real-world citation distributions, there were no such difficulties in the calculation of the top 10% thresholds reported in Table 1.

[2] The approach proposed by Leydesdorff et al. (2011) builds on an earlier paper by Bornmann and Mutz (2011). Some variants of and alternatives to the approach of Leydesdorff et al. are discussed by Leydesdorff and Bornmann (2011, in press), Rousseau (2011, 2012), and Schreiber (in press). We will discuss the proposals of Rousseau and Schreiber later on in this section.

We now consider the approach proposed by Pudovkin and Garfield (2009). Like in the approach of Leydesdorff et al. (2011), a percentile is determined for each publication. However, the way in which this is done is different. In the case of multiple publications with the same number of citations, Pudovkin and Garfield put the tied publications in a random order and calculate their average percentile. This average percentile is then attributed to all the tied publications. In our example, the publications with 10 citations have percentiles 91 / 105 = 86.7, 92 / 105 = 87.6, ..., 100 / 105 = 95.2, which yields an average percentile of 91.0. Because 91.0 ≥ 90, the publications with 10 citations are counted as belonging to the top 10% of their field, and we end up with 15 / 105 = 14.29% top 10% publications.

Another approach is taken in the Scimago Institutions Rankings (Bornmann et al., 2012). In this approach, the set of top 10% publications is defined in such a way that it always includes at least 10% of the publications in a field. This means that in our example the publications with 10 citations are counted as top 10% publications. Hence, like in the approach of Pudovkin and Garfield (2009), we have 15 / 105 = 14.29% top 10% publications. The approach taken in the Scimago Institutions Rankings is equivalent to the approach proposed by Rousseau (2011, 2012).

In the *Science and Engineering Indicators* report of the National Science Board (2012), the approach that is taken is exactly opposite to the approach of the Scimago Institutions Rankings. In the National Science Board approach, the set of top 10% publications is defined in such a way that it always includes at most 10% of the publications in a field. Clearly, in our example, this approach yields 5 / 105 = 4.76% top 10% publications. The National Science Board approach is equivalent to the approach of Leydesdorff et al. (2011) discussed above.

All approaches discussed until now fail to produce exactly 10% top 10% publications. Moreover, using these approaches, the degree to which top 10% publications are over- or underrepresented is likely to differ across fields and over time. This decreases the accuracy of inter-field and intertemporal comparisons. We now discuss two approaches that address this problem.

Recently, one of us proposed an approach that essentially boils down to abandoning the binary distinction between publications that belong to the top 10% of their field and publications that do not belong to the top 10% (Schreiber, in press). In the proposed approach, publications may belong to the top 10% with a certain fraction. Like in the approach of Pudovkin and Garfield (2009), publications with the same number of citations are put in a random order. However, unlike Pudovkin and Garfield's approach, no average percentile is calculated for tied publications. Instead, the following three steps are taken:
1. For each tied publication, a percentile is calculated. This is done based on the (random) order in which the tied publications were put.
2. Based on each publication's percentile, the proportion of the tied publications belonging to the top 10% of their field is calculated.
3. Each tied publication is counted as belonging to the top 10% of its field with a fraction equal to the proportion calculated in step 2.

We use our example to illustrate this approach. The publications with 10 citations have percentiles 90 / 105 = 85.7, 91 / 105 = 86.7, ..., 99 / 105 = 94.3.[3] It follows that 5

---
[3] Schreiber (in press) calculates the percentile of a publication in a slightly different way than Pudovkin and Garfield (2009). In the calculation of the percentile of a publication, Pudovkin and Garfield include

of the 10 publications belong to the top 10%, yielding a proportion of 5 / 10 = 0.5. Hence, a fraction of 0.5 of each of the publications with 10 citations is counted as belonging to the top 10%. Taking into account the 5 publications with 20 citations, this results in (0.5 × 10 + 5) / 105 = 9.52% top 10% publications.

As the example makes clear, the approach proposed by Schreiber (in press) comes closer to having 10% top 10% publications. Nevertheless, Schreiber's approach still does not produce exactly the right proportion of top 10% publications.[4]

An alternative approach has been in use for a number of years at the institute with which one of us is affiliated, the Centre for Science and Technology Studies of Leiden University. This approach is briefly explained by Van Leeuwen, Visser, Moed, Nederhof, and Van Raan (2003). Similar approaches are used by Colliander and Ahlgren (2011) and Lewison et al. (2007).

In this approach, a threshold is chosen and all publications whose number of citations is at least equal to the threshold are considered to belong to the top 10% of their field. The threshold is chosen in such a way that the over- or underrepresentation of top 10% publications is as small as possible. Nevertheless, there will usually be an over- or underrepresentation. A normalization is used to correct for this. To illustrate the normalization, we again use our example. The smallest over- or underrepresentation of top 10% publications is obtained by choosing a threshold of 10 citations. There are 15 publications that meet this threshold, resulting in 15 / 105 = 14.29% top 10% publications.[5] This means that there are 14.29% / 10% = 1.429 times as many top 10% publications as there should be. We therefore need a normalization factor of 1 / 1.429 = 0.700. Let us now consider a research group that has 9 publications without citations and 1 publication with 20 citations. Without the normalization, this research group has 1 / 10 = 10% top 10% publications. Hence, the conclusion would be that in terms of top 10% publications the group has exactly an average performance. However, using the normalization, the group has only 0.700 × 10% = 7.00% top 10% publications, which shows that the group in fact has a below average performance. The difference between the outcomes obtained with and without the normalization is due to the overrepresentation of top 10% publications. As a consequence of this overrepresentation, the outcome obtained without the normalization is too positive. We note that the normalization is defined in such a way that applying it to the field as a whole yields exactly 10% top 10% publications.

The approach discussed above produces exactly 10% top 10% publications, thereby allowing for accurate inter-field and intertemporal comparisons. However, the approach has another property that we consider less attractive. To see this, suppose that in our example one of the publications with 10 citations is replaced by a publication with 9 citations. The threshold then remains at 10 citations, but instead of 15 there are only 14 publications that meet the threshold. This results in 14 / 105 = 13.33% top 10% publications and, consequently, a normalization factor of 10% / (14 / 105) = 10% / 13.33% = 0.750. Using this normalization factor, the research group

---

the publication of interest in the numerator. Schreiber does not include this publication in the numerator.

[4] Including the publication of interest in the numerator of the percentile calculation, as is done by Pudovkin and Garfield (2009), would not solve this problem. The percentiles would range from 91 / 105 = 86.7 to 100 / 105 = 95.2, so that 6 publications with 10 citations would belong to the top 10%, resulting in (0.6 × 10 + 5) / 105 = 10.48% top 10% publications. Hence, we again do not have exactly the right proportion of top 10% publications.

[5] Setting the threshold at 11 citations would yield only 5 / 105 = 4.76% top 10% publications, which is further away from the desired 10% than the 14.29% obtained using a threshold of 10 citations.

considered above has 0.750 × 10% = 7.50% top 10% publications, an increase of 0.50% compared with the original situation. What we consider counterintuitive is that changing the number of publications that are exactly at the threshold affects a research group that itself does not have any publications at the threshold. In other words, the 'value' of a research group's frequently cited publications may increase or decrease as a consequence of changes that take place elsewhere in the citation distribution.

We have a similar but more extreme situation if instead of replacing a publication with 10 citations by a publication with 9 citations there is a change in the other direction. Suppose that of the 10 publications with 10 citations 2 are replaced by publications with 11 citations. The threshold then increases from 10 to 11 citations, and the number of publications that meet the threshold decreases from 15 to 7, yielding 7 / 105 = 6.67% top 10% publications.[6] As a consequence, the normalization factor increases from 0.700 to 10% / (7 / 105) = 10% / 6.67% = 1.500. Using the increased normalization factor, the research group considered above has 1.500 × 10% = 15.00% top 10% publications. This percentage is more than twice as high as in the original situation, in which the group had only 7.00% top 10% publications. Paradoxically, the research group has benefited from an increase in the number of citations to publications that are not its own. This is clearly counterintuitive.

## 4. Alternative approach to calculating percentile-based indicators

We now discuss an alternative approach to calculating percentile-based indicators. This approach produces exactly 10% top 10% publications and does not have the less attractive property discussed at the end of the previous section.[7] Like the approach of Schreiber (in press), our alternative approach counts publications as belonging to the top 10% of their field with a certain fraction. However, the way in which this fraction is determined is different from Schreiber's approach. We note that the alternative approach is already in use in the most recent edition of the Leiden Ranking, a university ranking based on bibliometric indicators that is produced by the Centre for Science and Technology Studies of Leiden University (Waltman et al., in press; www.leidenranking.com).

We again consider our example of a field with 105 publications: 90 without citations, 10 with 10 citations each, and 5 with 20 citations each. Using this example, our approach can be explained as follows. Each of the 105 publications represents 1 / 105 = 0.952% of the citation distribution of the field, as visualized in the third row of Figure 1. Hence, together the 90 publications without citations represent 90 × 0.952% = 85.71% of the citation distribution, the 10 publications with 10 citations represent 10 × 0.952% = 9.52% of the citation distribution, and the 5 publications with 20 citations represent 5 × 0.952% = 4.76% of the citation distribution. This is indicated in the fourth row of Figure 1. We are interested in the top 10% of the citation distribution, highlighted in the first row of Figure 1. Clearly, the 5 publications with 20 citations belong to the top 10%, while the 90 publications without citations do not. What about the 10 publications with 10 citations? These publications are assigned fractionally to the top 10% in such a way that we end up with exactly 10% top 10% publications. The 5 publications with 20 citations cover almost half (4.76%) of the top 10% of the citation distribution. The other half (5.24%) needs to be covered by the 10 publications with 10 citations. To accomplish this, the segment of the citation distribution covered by publications with 10 citations needs to be split into two parts, one part covering 5.24% of the distribution, the other part covering the remaining 9.52% – 5.24% = 4.29%. The part that covers 5.24% of the distribution is then considered to belong to the top 10% of the distribution. The other part belongs to the bottom 90% of the distribution. Splitting the segment of the distribution covered by publications with 10 citations is done by assigning each of the 10 publications to the top 10% with a fraction of 5.24% / 9.52% = 0.550. In this way, we obtain exactly 10% top 10% publications, since $(0.550 \times 10 + 5) / 105$ equals exactly 10%.

---

[6] Keeping the threshold at 10 citations would yield 15 / 105 = 14.29% top 10% publications, which is further away from the desired 10% than the 6.67% obtained using a threshold of 11 citations.

[7] This alternative approach was already briefly mentioned by Schreiber (in press), but Schreiber did not investigate it further. See also the reply to Schreiber by Leydesdorff (in press).

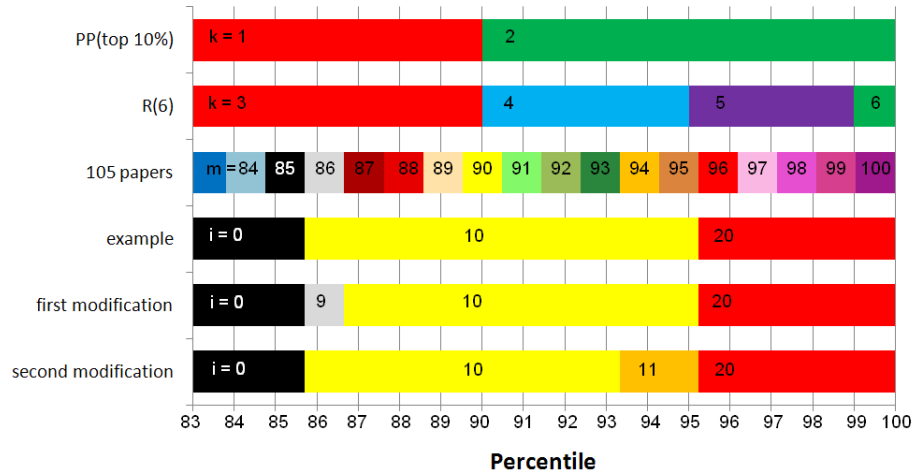

Figure 1. Percentile intervals $k$ for the $PP_{top\ 10\%}$ indicator (first row) and for the R(6) indicator of Leydesdorff et al. (2011) (second row) and publication intervals $m$ for the 105 publications in our example. First each publication is shown separately (third row), then intervals comprising publications with the same number of citations $i$ are shown (fourth row), and finally the two scenarios discussed in the text are indicated: The scenario in which 1 publication has 9 rather than 10 citations (fifth row), and the scenario in which 2 publications have 11 rather than 10 citations (sixth row). Vertical lines indicate the boundaries of the standard 100 percentile intervals. Note that only the upper end of the percentile scale is plotted, since the lower part is not interesting for the discussion. So in the second row only 4 of the 6 percentile intervals are shown, and in the third row only 18 of the 105 publications are indicated.

To illustrate the calculation of the $PP_{top\ 10\%}$ indicator using our above-discussed approach, let us first consider the same research group as we did at the end of the previous section. This group has 9 publications without citations and 1 publication with 20 citations. Publications with 20 citations fully belong to the top 10% of the citation distribution, and therefore the group has 1 / 10 = 10% top 10% publications. Suppose now that there is a second research group which has 9 publications without citations and 1 publication with 10 citations. Because publications with 10 citations belong to the top 10% of the citation distribution with a fraction of 0.550, this group has 0.550 / 10 = 5.50% top 10% publications.

Like in the previous section, suppose now that in our example one of the 10 publications with 10 citations is replaced by a publication with 9 citations. This scenario is visualized in the fifth row of Figure 1. Following the calculations shown above, we find that the fraction with which the remaining 9 publications with 10 citations are assigned to the top 10% of the citation distribution increases from 0.550 to 5.24% / (9 × 0.952%) = 0.611. This increase does not affect our first research group, since this group does not have any publications with 10 citations. It does affect our second research group. The proportion of top 10% publications of this group increases from 5.50% to 6.11%.

Now consider the other scenario introduced in the previous section: Of the 10 publications with 10 citations, 2 are replaced by publications with 11 citations (as indicated in the last row of Figure 1). In this scenario, we have 7 publications with 11 or more citations, representing 7 × 0.952% = 6.67% of the citation distribution. Consequently, the 8 remaining publications with 10 citations are assigned to the top 10% with a fraction of (10% – 6.67%) / (8 × 0.952%) = 0.438. This is a decrease in comparison with the original situation, in which publications with 10 citations were assigned to the top 10% with a fraction of 0.550. This decrease does not affect our first research group. It affects only our second research group, whose proportion of top 10% publications decreases from 5.50% to 4.38%.

The above two scenarios illustrate a nice property of our approach to calculating percentile-based indicators: A (small) change in the number of citations of one or more publications that are exactly at the top 10% threshold affects only research groups that have publications at the threshold. Research groups without publications at the threshold are not affected. Contrary to the approach discussed at the end of the previous section, it is not possible for a research group to benefit from an increase in the number of citations to publications that are not its own.

## 5. Formal mathematical framework

Until now, our focus has been on one specific percentile-based indicator: The $PP_{top\ 10\%}$ indicator. In this section, we explain how our approach discussed in the previous section can be used more generally for any percentile-based indicator. To do so, we need a more formal mathematical framework.

Let $0 = p_0 < p_1 < ... < p_N = 1$ denote the boundaries of $N$ percentile intervals, where $N \geq 2$. The first interval is $[p_0, p_1]$, the second interval is $[p_1, p_2]$, etc. Let $s_1 < s_2 < ... < s_N$ denote the scores associated with the $N$ intervals. Informally, the calculation of a percentile-based indicator can be summarized as follows:

1. For each publication of for instance a research group, determine the percentile interval $[p_{k-1}, p_k]$ to which the publication belongs.
2. Determine the score $s_k$ of each publication based on the percentile interval to which the publication belongs.
3. Calculate the average score of all publications.

We emphasize that the above steps provide only an informal summary of the calculation of a percentile-based indicator. In the above steps, it is assumed that there is no ambiguity in determining the percentile interval to which a publication belongs. As we have seen earlier in this paper, this assumption often does not hold.

Table 2. Parameter values for the R(6) indicator of Leydesdorff et al. (2011).

| $k$ | 0 | 1 | 2 | 3 | 4 | 5 | 6 |
|---|---|---|---|---|---|---|---|
| $p_k$ | 0.00 | 0.50 | 0.75 | 0.90 | 0.95 | 0.99 | 1.00 |
| $s_k$ |  | 1 | 2 | 3 | 4 | 5 | 6 |

Before formally defining our approach to calculating percentile-based indicators, we briefly mention two special cases of our generic percentile-based indicator introduced above. The first special case is the $PP_{top\ 10\%}$ indicator. This indicator is obtained by setting $N = 2$, $p_1 = 0.9$, $s_1 = 0$, and $s_2 = 1$. The second special case is the R(6) indicator discussed by Leydesdorff et al. (2011). The R(6) indicator is obtained by setting $N = 6$ and by choosing the values of $p_0, p_1, ..., p_6$ and $s_1, s_2, ..., s_6$ listed in Table 2.

To formally define our approach to calculating percentile-based indicators, we need some additional mathematical notation. Let $c_i$ denote the number of publications in a particular field that have $i$ citations, and let $q_i$ denote the proportion of the publications that have fewer than $i$ citations. Hence, $q_i$ can be written as

$$q_i = \frac{\sum_{j=0}^{i-1} c_j}{\sum_{j=0}^{\infty} c_j}. \qquad (1)$$

Note that the length of the interval $[q_i, q_{i+1}]$ equals the proportion of the publications in a field with exactly $i$ citations.

We now consider the score of a publication with $i$ citations, denoted by $S_i$. In some cases, publications with $i$ citations fully belong to a single percentile interval. The score of a publication with $i$ citations then simply equals the score of the percentile interval. In other cases, however, publications with $i$ citations need to be fractionally assigned to two or more percentile intervals. In these cases, the score of a publication with $i$ citations equals a weighted average of the scores of the relevant percentile intervals. To implement this, $S_i$ is defined as

$$S_i = \sum_{k=1}^{N} \frac{O_{ik}}{q_{i+1} - q_i} s_k, \qquad (2)$$

where $O_{ik}$ denotes the overlap of the intervals $[p_{k-1}, p_k]$ and $[q_i, q_{i+1}]$, or more formally,

$$O_{ik} = \max(\min(p_k, q_{i+1}) - \max(p_{k-1}, q_i), 0). \qquad (3)$$

If $c_i = 0$, the numerator and the denominator in (2) both equal zero. In this case, $S_i$ is not defined. The fractions $O_{ik} / (q_{i+1} - q_i)$ in (2) are the fractions with which a publication is assigned to different percentile intervals. These are the fractions that we discussed in the previous section. If there is a $k$ for which $[q_i, q_{i+1}]$ is enclosed by $[p_{k-1}, p_k]$, then $O_{ik} / (q_{i+1} - q_i) = 1$. It then follows from (2) that $S_i = s_k$. If there is no $k$ for which $[q_i, q_{i+1}]$ is enclosed by $[p_{k-1}, p_k]$, then $S_i$ equals a weighted average of two or more scores $s_1, s_2, ..., s_N$, where the weights are given by the fractions $O_{ik} / (q_{i+1} - q_i)$.

Given the scores of publications, the calculation of a percentile-based indicator is straightforward. Let $n_i$ denote the number of publications of a research group that have $i$ citations. A percentile-based indicator for the research group is calculated as

$$\text{PBI} = \frac{\sum_{i=0}^{\infty} n_i S_i}{\sum_{i=0}^{\infty} n_i}. \qquad (4)$$

Hence, a percentile-based indicator simply equals the average score of the publications of the research group.[8]

What remains to be done is to show that percentile-based indicators calculated using (4) behave properly. More specifically, we want a percentile-based indicator not to have any biases in favor of or against particular fields. In other words, when calculating a percentile-based indicator for all publications in a field taken together, the same outcome should be obtained for each field, irrespective of the particular citation distribution by which a field is characterized.

Mathematically, to show that percentile-based indicators calculated using (4) behave properly in the above-defined sense, we need to set $n_i$ equal to $c_i$ in (4) and we need to prove that the outcome of (4) does not depend on the citation distribution given by $c_0, c_1, \ldots$. By substituting (1) into the denominator of (2), substituting (2) into (4), and setting $n_i = c_i$ in (4), we obtain after some simplification and rewriting

$$\text{PBI} = \sum_{k=1}^{N} \sum_{i=0}^{\infty} O_{ik} s_k \ . \tag{5}$$

It is not difficult to see that

$$\sum_{i=0}^{\infty} O_{ik} = p_k - p_{k-1} . \tag{6}$$

In other words, the combined overlap of the interval $[p_{k-1}, p_k]$ with the intervals $[q_0, q_1], [q_1, q_2], \ldots$ equals the length of the interval $[p_{k-1}, p_k]$. Substituting (6) into (5) yields

$$\text{PBI} = \sum_{k=1}^{N} (p_k - p_{k-1}) s_k \ . \tag{7}$$

Eq. (7) does not depend on the citation distribution given by $c_0, c_1, \ldots$. It only depends on the parameters $p_0, p_1, \ldots, p_N$ and $s_1, s_2, \ldots, s_N$. This proves that our approach to calculating percentile-based indicators leads to indicators that behave properly and that do not have any biases in favor of or against particular fields. Because of this, we consider our approach preferable over other approaches for which (7) does not hold, like most of the approaches discussed in Section 3.

---

[8] To avoid additional technicalities, it is assumed in (4) that all publications of the research group belong to the same field. Dropping this assumption is straightforward.